\documentclass[twocolumn,aps,prb]{revtex4}
\usepackage{graphicx}
\usepackage{color}
\usepackage{amsmath}
\usepackage{hyperref}

\hypersetup{
	colorlinks = true,
	linkcolor = blue,
	citecolor = blue
}

\newcommand{\be}{\begin{equation}}
\newcommand{\ee}{\end{equation}}

\newcommand{\bea}{\begin{eqnarray}}
\newcommand{\eea}{\end{eqnarray}}

\newcommand{\lb}{\left[}
\newcommand{\rb}{\right]}
\newcommand{\lp}{\left(}
\newcommand{\rp}{\right)}

\renewcommand{\Re}{{\rm \, Re\,}}

\renewcommand{\vec}[1]{{\boldsymbol #1}}
\renewcommand{\epsilon}{\varepsilon}

\begin{document}
\title{Three-Body Bound States of Quantum Particles: Higher Stability Through Braiding} 
\author{Sophie Fisher, Olumakinde Ogunnaike, Leonid Levitov}
\affiliation{Massachusetts Institute of Technology, Cambridge, Massachusetts 02139, USA}
\date{\today}

\begin{abstract}

Cold atoms embedded in a degenerate Fermi system interact via a fermionic analog of the Casimir force, which is an attraction of a $-1/r$ form at distances shorter than the Fermi wavelength. 
Interestingly, the hydrogenic two-body bound states do not form in this regime because the interaction strength is too weak under realistic conditions, and yet the three-body bound states can have a considerably higher degree of stability. As a result, the trimer bound states can form even when the dimer states are unstable. 
A quasiclassical analysis of quantum states supported by periodic orbits singles out the ``figure-eight'' orbits, predicting bound states that are more stable than the ones originating from circular orbits. The discrete energies of these states form families of resonances with a distinct structure, enabling a direct observation of signatures of figure-eight braiding dynamics. 
\end{abstract}
\maketitle

Cold atom systems provide a unique platform to investigate surprising properties of quantum few-body states. One of the remarkable predictions of quantum theory is that when the interactions between particles are not strong enough to support a two-body bound state, they may nonetheless support three-body bound states. A celebrated example of this behavior is Efimov trimers formed by particles interacting through short-range attractive interactions that are nearly resonant\cite{efimov1970,naidon2017}. In this case, an infinite tower of three-body bound states forms even though two-body bound states are unstable. These trimer states have a peculiar nested shell structure related to ``discrete scale invariance'' and limit cycles in the renormalization group\cite{leclair2004,bedaque1999,wilson2002}. Efimov states have been a focus of active research in nuclear and cold atom physics, culminating in recent observations of a hierarchy of these states\cite{huang2014,tung2014,pires2014}.

Here we introduce a qualitatively different kind of bound states. We discuss bound states of two and three bosons embedded in a degenerate Fermi sea of cold atoms. In this case, a fermion-mediated interaction arises through the so-called RKKY mechanism\cite{ruderman_kittel_1956,kasuya_1956,yosida_1957}, 
giving
\be\label{eq:RKKY}
U(R)= -3\alpha  \frac{\sin (2k_F R)-2k_F R\cos (2k_F R)}{R^4}
,
\ee
where $k_F$ is the Fermi momentum, and the interaction strength $\alpha$ depends on the boson-fermion scattering length and particle masses as discussed below. The fermion-mediated RKKY interaction between bosonic atoms was demonstrated in recent experiments with a Bose-Einstein condensate of Cs atoms embedded in a degenerate Fermi gas of Li atoms\cite{chin2019}. As we will see, the bound states formed due to the RKKY interactions originate from classical orbit quantization and, therefore, form families of resonances with distinct structure.

\begin{figure}[b]
\includegraphics[width=0.99\linewidth]{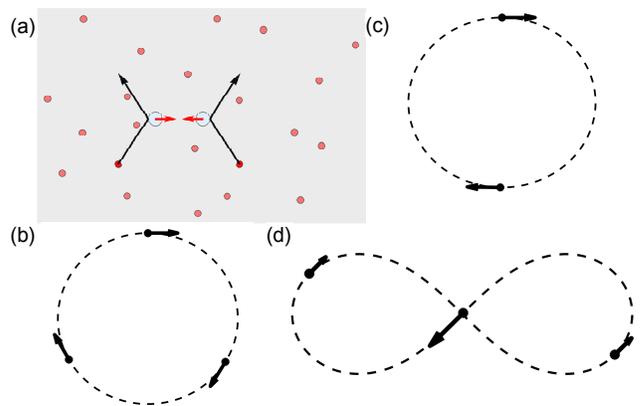}
\caption{(a) Schematic of attractive force between bosons (blue), mediated by the ``inward pressure" due to scattering of fermions (red). (b) A two-body orbit. (c) A circular three-body orbit. (d) A figure-eight three-body orbit. }
\label{fig1}
\end{figure}

In this case, in place of the short-range Efimov interaction, particles 
interact via a long-range interaction. Crucially, this interaction 
is an attractive power law in $R$ in a relatively wide range of 
distances:
\be
U(R)=-\frac{\alpha}{R}
,\quad 2k_F R < 1
.
\ee
The origin of attraction can be seen physically: an anisotropy in the flux of fermions incident on the two bosonic atoms, with one atom shielding the other one, causes a net attractive force between the atoms via the fermionic Casimir effect.\cite{mele_2008} This is illustrated in Fig.\ref{fig1}(a). 

Naturally, the behavior $U(R)\sim -1/R$ prompts the question of whether the two-particle bound states of a hydrogenic type, pictured schematically in Fig.\ref{fig1}(b), can occur. Nominally, the stability of such hydrogenic states would not depend on the interaction strength, $\alpha$. However, since the radius of a hydrogenic states scales inversely with interaction strength, the $\alpha$ values do determine for which states $2k_F R > 1$; at larger $R$, the RKKY interaction falls off too quickly to support bound states. Our analysis indicates that, while $\alpha$ can be tuned by varying the fermion density and the boson-fermion scattering length, the RKKY interactions is too weak to support two-body bound states. Surprisingly, however, the requirements for the formation of three-body bound states are not as stringent and such states can indeed occur under realistic conditions. 

The stability of these three-body states stands in sharp contrast to that of the classical three-body problem. Classically, gravitational three-body orbits are prone to decay into the more stable configurations of a two-body orbit and an unbound third body. However, this decay channel is quenched in the absence of two-body bound states. Strikingly, this makes the quantum three-body ``orbits'' far more stable than the classical three-body orbits: our trimer states are effectively stabilized by the lack of two-body bound states (a behavior similar to that of Efimov states).

Further, as we will see, the three-body states originating from the RKKY interaction are highly sensitive to orbit geometry. As appropriate for bound states supported by a long-range attraction, the underlying physics here can be best understood in a quasiclassical framework. Below, we apply quasiclassical quantization using the Gutzwiller trace formula framework. We consider, as primary examples, two simple periodic orbits of the three-body problem: the circular and the figure-eight orbits\cite{moore1993} pictured in Fig.\ref{fig1} (c) and (d), respectively. For both orbits the dynamics are locally stable, such that small perturbations remain small at all times.

These orbits  share similar dynamic symmetry.  The figure-eight orbit is a celebrated solution to the planar three-body problem in which three equal-mass particles travel around {\it the same figure-eight curve} with time shifts equal to $1/3$ of the period, as illustrated in Fig.\ref{fig1}(d).

The figure-eight is the simplest periodic orbit in a large family discovered by Moore\cite{moore1993}. Originally it was located numerically using a functional gradient descent procedure described in the Supplementary Material\cite{Supplemental}; its existence was later confirmed by a rigorous analysis\cite{Chenciner2000}. Circular orbits on which two and three particles are chasing each other with $1/2$ and $1/3$ period time shifts, as shown in Fig.\ref{fig1}(b,c), provide a natural comparison to figure-eight orbits. 

Our analysis of the quantum states associated with these orbits predicts Rydberg-like energy spectra

\be\label{eq: E_n}
E_n=-C\frac{\alpha^2 m}{4\hbar^2 n^2}
,
\ee
where the prefactor $C$ depends on the orbit geometry and $n$ is a positive integer with an upper bound where our approximation breaks down: $1 \le n \le n_{max}$. For distinguishable particles $n$ takes all positive integer values, whereas for identical bosonic particles $n=1+3k$. The energy spectrum, Eq.\eqref{eq: E_n}, is written in a form that facilitates comparison with the conventional Rydberg spectrum. For two particles of equal masses, the latter is given by setting $C=1$.  For the three-body circular-orbit dynamics we find  $C=18$; for the figure-eight orbit we find $C\approx 34$. The large $C$ values indicate that the three-body states are considerably more stable than the two-body states. 

The difference in binding strengths can be traced to particles' orbit geometry and topology. Three particles moving along a circular orbit will clearly interact more strongly than two particles on the same orbit. Stronger binding for the figure-eight states as compared to the circular-orbit states can be attributed to the peculiar figure-eight braiding dynamics. The dynamics is such that the three particles come much closer to each other than particles moving along a circular orbit of comparable orbit size, which translates into a larger binding energy and higher stability. 

These braiding dynamics should be contrasted with the 
schemes for 
braiding of particles with anyon statistics\cite{Anyons, Honeycomb} 
or the physical braiding of qbits represented by trapped ions or the like to perform actions of quantum gates\cite{surface_code1,surface_code2,Ion_Trap,Phase_Gate}.  These operations typically require the intervention of an external source for confinement or control of the dynamics, while the braiding described here may exist in an isolated system.  The braiding described below illustrates the largely unexplored quantum-coherent phenomena for this RKKY interaction.

Further, we note that the simple hydrogenic $-1/R$ model for the fermion-mediated attraction, assumed in Eq.\eqref{eq: E_n}, is perfectly sufficient for assessing stability of low-lying states, a question that will be the focus of this paper. The RKKY interaction deviates from the simple model outside the range $2k_F R\ll1$, where it falls off more rapidly than $1/R$. 
This behavior does not matter for the low-lying states so long as their localization radius is small enough: $2k_FR\ll1$; yet, it is detrimental for the high-$n$ states from Eq.\eqref{eq: E_n} because a larger $n$ translates into a larger orbit radius. For realistic $\alpha$ values, which are relatively small, we find that the spatial extent of \textit{all} hydrogenic two-body states exceeds the Fermi wavelength, i.e. they fall outside the range $2k_FR\ll1$ in which the $-1/R$ form holds. In contrast, for three-body states the value of $\alpha$ is high enough to push the orbit radius under the $2k_FR\ll1$ bound for a finite number of low-lying states, ensuring their stability.

To estimate the realistic interaction strength we use the parameter values from the recent experiment in which the fermion-mediated interaction was observed\cite{chin2019}. As shown gives rise to a $-1/R$ potential with which we can analyze bound states for two and three bosonic particles

\be \label{alphavalue}
U(R) = - \frac{g_{BF}^2 m_F k_F^3}{3 \hbar^2 \pi^3 R} = -\frac{\alpha}{R}
,\quad
\alpha= 0.0784 \frac{ \hbar^2  a_{BF}^2 k_F^3}{m_p} 
\ee
where $m_F$ and $m_p$ are the masses of the fermion and proton, $g_{BF}$ and $a_{BF}$ are the boson-fermion coupling and scattering length, and $k_F$ is the Fermi momentum.

It is instructive to start our discussion of the spectra with examining the two-body ground states. 
  We estimate the typical boson separation by considering the Bohr's radius, $R_0$ with reduced mass, $\frac{m_B}{2}$: 
\be\label{eq:2k_F R_0}
2k_F R_0 = \frac{4 \hbar^2 k_F}{m_B \alpha} =\frac{4\cdot a_{BF}^{-2} k_F^{-2} }{133 \cdot 0.0784}= 1.39 \cdot 10^7\lp \frac{a_0}{a_{BF}}\rp^2
\lp \frac{k_F^{(0)}}{k_F} \rp^2.
\ee
Here, motivated by the form of $\alpha$ in Eq.\eqref{alphavalue}, we have chosen to normalize $a_{BF}$ and $k_F$ by Bohr's radius $a_0$ and $k_F^{(0)}\approx\pi\,{\rm \mu m^{-1}}$, the value from Ref.\onlinecite{chin2019}. 

Which parameter values can support states with radii in the range where our hydrogenic approximation holds?  Choosing the realistic values 
$a_{BF} = 100a_0$ and $ \frac{k_F}{k_F^{(0)}} = 10$ we find
\be
2k_FR_0 \approx 14
.
\ee 
This value is too large to justify the approximation $2k_FR\ll1$, indicating that no two-body bound states occur in this case. However, the conditions for confinement can be relaxed by tuning system parameters. Choosing higher values $a_{BF}=500a_0$ and $k_F = 20 k_F^{(0)}$, we see 
\be
2k_FR_0 = 0.14
,
\ee 
which is justifiably small. However, in this limit, our hydrogenic approximation is likely to be invalid for other reasons.  Because the orbit size is smaller than the scattering length, $R<a_{BF}$,  
the bosons come close enough to form a many-body \textit{condensate} in place of approximately isolated few-body states. 
In addition, accessing such high values of $a_{BF}$ and $k_F$ may be practically challenging.  Further discussion and spectral estimates for accessible three-body states may be found in the Supplemental Material\cite{Supplemental}.

Next we discuss three-body bound states and argue that the conditions for supporting these states are less stringent than those for two-body states. The three-body states are described by the Hamiltonian 
\be\label{three_hamiltonian}
H(\vec r_i,\vec p_i)=\sum_{i}\frac{\vec p_i^2}{2m}+\sum_{i\ne i'}U(\vec r_i-\vec r_{i'})
,\quad
i,i'=1,2,3
.
\ee
We consider the long-range interactions $U(r) \sim -1/r$ and employ quasiclassical methods, wherein bound states arise from quantized periodic orbits. Importantly, unlike the two-body problem, the three-body problem is non-integrable; as a result, the dynamics is chaotic in most of its phase space. Yet, islands of stability associated with certain periodic orbits are known to exist, giving rise to families of discrete states. 

We analyze the three-body bound states using Gutzwiller's semiclassical quantization of non-integrable Hamiltonian systems\cite{gutzwiller1971}. Gutzwiller's approach identifies the contribution to the density of states from quantum states associated with periodic orbits, which allows one to separate discrete states from the chaotic continuum. We apply this approach to the circular and figure-eight orbits pictured in Fig.\ref{fig1} (c) and (d). We first consider distinguishable particles; in this case the period of the orbit equals the time it takes each particle to undergo a full revolution. We then consider the case of indistinguishable particles. In this case, the period is reduced by $1/3$, since the particles reach a permuted version of the initial point in phase space after a third of the period, and thus arrive at the same quantum state.

The Gutzwiller trace formula approximates the density of states of a non-integrable Hamiltonian system as\cite{gutzwiller1971}

\begin{equation}
	D(E) = \bar{D}(E) + \Re \sum\limits_{p} \frac{T_p}{\pi \hbar} \sum\limits_{r = 1}^{\infty} A_{p,r} e^{ \frac{irS_p }{\hbar} - \frac{i\sigma_{pr} \pi}{2} }
	\label{traceformula}
\end{equation}
where $p$ sums over all primitive (non-repeated) periodic orbits with energy $E$, period $T_p$, action $S_p = \int \vec p \cdot d \vec q$, and $r$ sums over all repetitions of a primitive orbit. Here $\sigma_{pr}$ is the Maslov index for the $r$-th repetition of the primitive orbit $p$, see Ref.\onlinecite{creagh1990}. The amplitude factor $A_{p,r} = |\det (M_p^r  - 1) |^{-1/2}$ is a function of the stability matrix, 
$M_p$ that describes the local flow linearized about the primitive orbit $p$. The quantity $\bar{D}(E)$ is the average density of states of the system which depends smoothly on energy (the Thomas-Fermi contribution associated with the chaotic states). In our calculations, we will disregard this term because we only care about the oscillatory contribution to the density of states
coming from the sum over classical periodic orbits. For simplicity, we set the amplitude factor $A_{p,r} = 1$ and also assume the Maslov index to be additive over successive repetitions $r$ of a primitive orbit $p$; denoting  the index for one revolution of the 
orbit as $\mu\equiv \sigma_{p1}$ we write $\sigma_{pr} = r \mu$. The oscillatory contribution to the density of states is then given by a sum of terms multiplicative in $r$:
\begin{equation}
	\delta D(E) = \Re \sum\limits_{p} \frac{T_p}{\pi \hbar} \sum\limits_{r = 1}^{\infty} \exp\lb i r \lp\frac{S_p}{\hbar} - \frac{\mu \pi}{2} \rp \rb .
	\label{traceformula2}
\end{equation}
The validity of the simplifying assumptions that lead to Eq.\eqref{traceformula2} 
will be discussed in detail elsewhere.

An essential property of periodic orbits allowing them to support discrete states is linear stability. While generally rare for periodic orbits of the three-body problem, this property holds for the orbits of interest. The stability of the three-body circular orbit is well known\cite{moore1993}; 
for the figure-eight orbits it 
was demonstrated in Refs.\onlinecite{Kapela_2007,roberts_2007}
by verifying that all eigenvalues of the stability matrix lie on the unit circle, and later proven 
rigorously 
in Ref. \onlinecite{Stability1}.

Another important aspect of our three-body problem is that orbits are 
unique up to symmetries of the equations of motion, including translation, rotation, and rescaling of the $\vec r_i$ and $\vec p_i$ variables
\be\label{scaling_symmetry}
\vec r_i\to \frac1{\beta}\vec r_i
,\quad
\vec p_i\to \beta^{1/2}\vec p_i
,\quad
H\to \beta H
.
\ee

We use this scaling symmetry to re-write our trace formula in terms of a single``reference" orbit. Indeed, Eq.\eqref{scaling_symmetry} defines a continuous family of orbits that are equivalent up to a rescaling. If $\vec r(t)$, $\vec p(t)$ define a 
solution with energy $E$, period $T$, and action $S$, then another solution is given by 
\be
\vec r'(t) = \beta^{-1}\vec r(\beta^{3/2} t)
,\quad
\vec p'(t) = \beta^{1/2}\vec p(\beta^{3/2} t)
\ee
with energy $E' = \beta E$, period $T' = \beta^{-3/2} T$ and action $S' = \beta^{-1/2} S$, for any $\beta > 0$. This scaling has consequences for our trace formula. Because each orbit with energy $E$ can contribute to the density of states only at $D(E)$, our calculations in Eq.\eqref{traceformula2} must incorporate the scaling relations. To proceed, we calculate the energy, period, and action of one particular ''reference'' 
orbit, which we label $\bar{E}$, $\bar{T}$ and $\bar{S}$. We define a scaling factor $\beta$ for an 
orbit with energy $E$, taken relative to an orbit with energy 
$\bar{E}$, such that $\beta_E=E/\bar{E}$. Then the action and the period of the rescaled orbit can be written as 
\be
S_E=
\beta_E^{-1/2} \bar{S}
,\quad 
T_E= 
\beta_E^{-3/2} \bar{T}
.
\ee 
Focusing on the case of distinguishable particles,
we can write the oscillatory contribution to the density of states from the entire family of figure-eight orbits: 
\begin{equation}
\delta D(E) = \Re \frac{\bar{T}\beta_E^{-3/2}}{\pi \hbar} 
 \sum\limits_{r = 1}^{\infty} \exp \lb i r \lp \frac{\bar{S}}{\hbar} 
\beta_E^{-1/2} - \frac{\mu \pi}{2}\rp \rb
.
\end{equation}
The sum over repetitions $r$ is a geometric series equal to 
\begin{align}
\delta D(E)  = \Re \frac{\bar{T}\beta_E^{-3/2}}{\pi \hbar} 
\frac{\exp\lb i \lp \frac{\bar{S}}{\hbar}
\beta_E^{-1/2} - \frac{\mu \pi}{2}\rp\rb }{1 - \exp\lb i \lp \frac{\bar{S}}{\hbar}
\beta_E^{-1/2} - \frac{\mu \pi}{2}\rp\rb} 
\label{poles}
\end{align}
%
From this, we can calculate a spectrum using a reference orbit as input; however the resulting spectrum should be independent of the reference orbit chosen.  The poles of Eq.\eqref{poles} lead to delta functions in the density of states whenever $\frac{\bar{S}}{\hbar}\big(\frac{\bar{E}}{E}\big)^{1/2} - \frac{\mu \pi}{2}  = 2\pi n$. Rearranging this condition, we find the energies $E_n$ of 
an orbit with distinguishable particles, labeled by a quantum number $n$:
\begin{equation}
E_n = \frac{\bar{E} \bar{S}^2}{4\hbar^2\pi^2(n + \frac{\mu}{4})^2}.
\label{energies}
\end{equation}
Further focusing on the figure-eight solution and using the numerical method described in 
the Supplementary Material\cite{Supplemental}, we find $\bar{E} = -1.2935 \alpha$ and $\bar{S} = 16.1609 (\alpha m)^{1/2}$. The values $\bar{E}$ and $\bar{S}$ depend on the orbit used as an initial condition in the relaxation dynamics. However, their product $\bar{E} \bar{S}^2$ is a universal constant independent of the details of the procedure. Comparing to Eq.\eqref{eq: E_n}, we evaluate 
\be
C=\bar{E} \bar{S}^2/\pi^2 \alpha^2m=34.23...
,
\ee
and 
set the Maslov index to its one-body and two-body value $\mu=4$ (and shift $n+1\to n$) to yield our expected spectrum.



Our analysis of the figure-eight orbit can be easily extended to circular orbits. In the case of distinguishable particles, we need only calculate $\bar{E}$ and $\bar{S}$ for a reference radius, $r = 1$ in appropriate units, 
and plug these values into Eq.\eqref{energies}.  
We find values
\be
\bar{E} = -\frac{3^{1/2} \alpha}{2}
,\quad
\bar{S} = 2\pi  (3^{3/2}\alpha m)^{1/2}
,
\ee 
which gives $E_n = -9 \alpha^2 m/2\hbar^2(n + \frac{\mu}{4})^2$, which is nothing but Eq.\eqref{eq: E_n} with $C=18$. 



We now consider how Eq.\eqref{energies} 
must be modified for the case of indistinguishable particles. This is done by accounting for the permutation symmetry of the three-particle states. 
As noted above, the circular and figure-eight orbits share a dynamical symmetry.  This shared symmetry allows us to consider either the circular orbit or figure-eight orbit with total period $T$ and action $S$, where the particles start at the initial point $(\vec r,\vec p)$ in phase space. After a time $T/3$, the particles reach a permuted version of the initial phase space point, $(P \vec r, P \vec p)$, where $P$ is the operator corresponding to the permutation $(123)$. Since the particles are indistinguishable, the system has reached a quantum state identical to the initial state. Thus, we can think of $T/3$ as the new period of the system. Then the action of the system becomes $S/3$. Since the Maslov index for the modified orbit is one third of the value for the original orbit, we replace $\bar{S}$ with $\bar{S}/3$ and $\mu$ with $\mu/3$ in Eq.\eqref{energies}. This yields the spectrum in Eq.\eqref{eq: E_n} with $n$ taking values $1$, $4$, $7$, $10$, and so on. 

A more intuitive way to arrive at this result is to consider the ground state for distinguishable particles for which the wavefunction is nodeless and is therefore identical to that for the ground state of bosonic particles. The number of nodes for the excited states, from permutation symmetry, must equal $3k$ for some positive integer $k$ value, which leads to Eq.\eqref{eq: E_n} with $n=1+3k$ as above. These selection rules for quantum numbers reflect the permutation symmetry of the three-body states of identical bosons. One can view this property as a special case of the constraints on the three-body states due to braiding dynamics of identical particles.

Lastly, as a consistency check, we verify that these states satisfy $2k_F R_0 \ll 1$.  Starting from the relation in Eq.\ref{eq:2k_F R_0} and rescaling it by the factors of $C=18$ and $C\approx 34$ for the two types of orbits gives
\be\label{eq:2k_F R_0_C}
2k_F R_0 =  \frac{1.39 \cdot 10^7}{C}\lp \frac{a_0}{a_{BF}}\rp^2
\lp \frac{k_F^{(0)}}{k_F} \rp^2.
\ee
Choosing the same values for the interspecies scattering length and the Fermi momentum as above, $a_{BF} = 100a_0$ and $ \frac{k_F}{k_F^{(0)}} = 10$, we find $2k_FR_0 \lesssim 1$ for the circular orbit and $2k_FR_0 < 1$ for the figure-eight orbit. The small values of $2k_FR_0$ justify our $-1/r$ approximation. Tighter confinement of the three-body states enables their higher stability over two-body states.  

We attribute the higher stability of the figure-eight states, as compared to the ones on circular orbits, to the ''intertwining'' character of the braiding dynamics that brings particles much closer together for the same orbit radius. The three-body states are further stabilized by the lack of stable two-body states, as discussed above.
We emphasize the general character of our analysis, which can be applied to both other three-body orbits and the $n$-body problem. This is because the derivation of Eq.\eqref{energies} depends only on the scaling properties of the system, and not on the number of particles. 

Interestingly, the figure-eight orbit is known to exist in the $n$-body problem for all odd $n \ge  3$ \cite{10.1007/978-3-0348-8268-2_6}, and one could thus use the same method to analyze the spectra of $n$-body bound states of the figure-eight orbit. As in the three-body case, quantum statistics of identical particles will manifest itself  through holes in the discrete spectrum. 

Quantum states associated with these orbits, if realized in experiment, can provide a unique opportunity to demonstrate braiding that results directly from unitary quantum evolution and does not depend on external driving. The potential for braiding in three-body dynamics and the sensitivity to the different resulting geometries make these states particularly promising systems for further study.

We are grateful to Alexander Turbiner, Chang Chin and Vladan Vuletic for useful discussions. 


\end{document}